# Use of Anodic TiO$_2$ Nanotube Layers as Mesoporous Scaffolds for Fabricating CH$_3$NH$_3$PbI$_3$ Perovskite-based Solid State Solar Cells

Raul Salazar,[1]† Marco Altomare,[1]† Kiyoung Lee,[1] Jyotsna Tripathy,[1] Robin Kirchgeorg,[1] Nhat Truong Nguyen,[1] Mohamed Mokhtar,[2] Abdelmohsen Alshehri,[2] Shaeel A. Al-Thabaiti,[2] and Patrik Schmuki[1,2]*

[1] Department of Materials Science and Engineering, WW4-LKO, University of Erlangen-Nuremberg, Martensstrasse 7, D-91058 Erlangen, Germany

[2] Department of Chemistry, King Abdulaziz University, Jeddah, Saudi Arabia

† These authors contributed equally

* Corresponding author: Tel.: +49-9131-852-7575, Fax: +49-9131-852-7582

Email: schmuki@ww.uni-erlangen.de

**A B S T R A C T**

We optimize the deposition of $CH_3NH_3PbI_3$ perovskite into mesoporous electrodes consisting of anodic $TiO_2$ nanotube layers. By a simple spin coating approach, complete filling of the tube scaffolds is obtained, that leads to interdigitated perovskite structures in conformal contact with the $TiO_2$ tube counterparts. Such assemblies can be used as solid state solar cells in hole-transporting material-free configuration, *i.e.*, the tube scaffold serves as electron collector and blocking layer, while the perovskite acts as visible light absorber and hole transporting material. We show that the complete filling of the tube scaffold with the perovskite is essential to improve the solar cell efficiency.

*Keywords:* $TiO_2$ nanotube electrode; anodization; perovskite; solid-state solar cell

1. **Introduction**

Significant improvements in the evolution of photovoltaic technology have been recently achieved by introducing perovskite-based materials in solar cell fabrication. In fact, the outstanding properties of these organo-metal halide light absorbers have, in a few years, led from low efficiency ($\eta=3.7\%$) short-life-operation DSSC-version devices to more robust solid-state solar cells that can reach power conversion efficiencies exceeding 16%[1–3] (perspective wise, cell efficiencies can in principle be pushed toward 20%[4,5]).

Key advantages of using a perovskite sensitizer are its efficient charge transport properties and light absorption ability,[5] and a well-developed solar cell geometry implies a planar thin film configuration in which the $TiO_2$ electrode typically consists of a submicrometer nanoparticle layer.[6]

However, the use of one-dimensional (1D) $TiO_2$ nanostructures (*e.g.*, nanorods, nanowires and nanotubes[7–9]) may be advantageous: with such electrodes it is possible for instance to deposit larger amounts of perovskite, this aiming at improving the light absorption ability. Moreover, 1D $TiO_2$ nanostructures are also promising candidates as electron collector scaffolds, owing to their directional charge transport properties.[9,10] Besides, a 1D $TiO_2$ electrode can still assume the role of "blocking layer", that is, it minimizes the occurrence of electrical short circuiting.[11,12]

Over the past few years, self-ordered anodic $TiO_2$ nanotube (NT) layers have attracted wide interest, due to their facile production and a wide range of control over their geometry.[13] In this work we discuss the optimization of the perovskite deposition into anodic $TiO_2$ nanotube scaffolds and study the feasibility of realizing hole transport material (HTM)-free perovskite-based solar cells: the perovskite acts as visible light absorber and hole transporting material.[14] In general, perovskite solar cells that are electrolyte-based show a loss of efficiency over time that is ascribed to the chemical instability of the perovskite material.[1,15] Therefore, as shown in Fig. 1(a), a solid-state configuration for the solar cells (*i.e.*, electrolyte-free) was adopted.

## 2. Results and discussion

Vertically aligned $TiO_2$ nanotube layers, as those shown in Fig. 1 and 2, with individual tube diameter of *ca.* 60-80 nm were grown by electrochemical anodization of Ti films (deposited on FTO slides) at 60V, in an $NH_4F$-containing ethylene glycol-based electrolyte (see Experimental section). Different thicknesses of the nanotube layers were fabricated by tuning the thickness of the evaporated Ti films on FTO, that is, metal films with thickness of *ca.* 0.2, 0.4 and 1 µm were completely anodized. The complete anodization of Ti films precedes the exposure of the underlying conductive FTO layer to the electrolyte that in turn would lead to sudden increase of the current density – just prior to this, the anodization experiments were terminated. This resulted in the formation of 0.4, 1 and 2 µm-thick $TiO_2$ nanotube layers. The increase of layer thickness (with respect to the Ti film thickness) is due to the volume expansion typically observed when anodically growing $TiO_2$ nanotube layers.[13,16]

In order to fabricate a functional solar cell geometry (such as in Fig. 1(a)), we found that a most relevant key is to optimize the perovskite deposition[17] approach towards complete filling of the $TiO_2$ tube scaffolds. Therefore, spin-coating deposition was investigated using perovskite solutions of different concentrations. Fig. 1(b)-(g) show 0.4 µm-thick $TiO_2$ nanotube layers that were coated with 10, 20 and 40 wt% $CH_3NH_3PbI_3$ perovskite solutions.

When a 10 wt% solution was used, the walls of the tubes could be decorated with isolated perovskite nanocrystals with rather broad size distribution, *i.e.*, in the 2-17 nm range (Fig. 1(b) and (c)). This morphology is typically observed when a classic mesoporous particle template is used.[15]

When depositing 20 wt% solutions, a nearly complete filling was obtained (Fig. 1(d)) – however, some voids were still left in the perovskite structure, this most likely due to solvent evaporation upon annealing (as outlined in the experimental section, the annealing step is required for crystallizing).

As clearly shown in Fig. 1(e)-(g), complete filling of the tubes was obtained by using 40 wt% solutions. In this case the perovskite material could be confined into the $TiO_2$ scaffolds and the 0.4 µm-thick tubes were fully filled, from the very bottom to the top, leading to perovskite morphology that resembles a replica of the tube scaffold. Noteworthy, we could finely cast the perovskite shape by using a single-step deposition method (and a tube layer template), which clearly simplifies the solar cell processing compared to previously reported multi-step solution-based approach.[18]

The effective deposition and crystallization of the perovskite was confirmed by means of X-ray diffractometry (see the XRD patterns of crystalline $TiO_2$ tube scaffolds before and after perovskite deposition in Fig. 1(h)).[15]

In order to complete the solar cell fabrication, we sputter-coated Au electrodes (100 nm-thick) onto the assemblies (Fig. 1(a)). J-V curves of the device were then measured under AM 1.5 simulated solar light irradiation, as shown in Fig. 2(a). Overall, solar cells fabricated by depositing the perovskite into the $TiO_2$ tube scaffolds from 10 wt% solutions showed no photoresponse. This can be explained by considering that the perovskite deposition from 10 wt% solutions leads only to isolated nano-sized crystals (*vide supra*) that do not provide a continuous medium for hole transport. Moreover, the ohmic behavior observed in the J-V curves is likely the result of a direct contact between the $TiO_2$ scaffold and the Au electrode. Thus, for an ideal perovskite filling and, at the same time, not to compromise the solar cell functionality, we found it important to prevent electrical shunts ascribed to direct contact between the electron collector and the Au electrode.

For this, not only we optimized the perovskite deposition to obtain complete filling of the tube layers (this aspect is discussed later) but we also extended the deposition of the perovskite to form a capping layer, of different thicknesses, over the tube scaffold (Fig. 2(b) and (c)). In principle, such a perovskite over-layer should be sufficiently thick to optimize the light absorption and prevent electrical shunts, but thin enough to grant effective hole transport towards the Au electrode.[2,19]

We found that a >500 nm-thick perovskite over-layer (see Fig. 2(b)) resulted in relatively low short circuit current density ($J_{sc}$) and efficiency ($\eta$) (Fig. 2(a)), this although such geometry shows excellent light absorbance (Fig. 2 (d)). In fact, this thick over-layer most likely limits the hole-transport and consequently leads to relatively low $J_{sc}$. Instead, thinner perovskite over-layers (<100 nm-thick, see Fig. 2(c)) yielded a much higher $J_{sc}$ of *ca.* 19 mA cm$^{-2}$, resulting in our highest cell efficiency of 5.0% (Fig. 2(a)).

Slightly higher efficiencies (*ca.* 6.5%) were recently reported for solar cells fabricated by depositing the perovskite absorber onto anodic $TiO_2$ nanotube membranes.[9] The scope of the work was not that of optimizing the light absorber deposition, and therefore only a limited perovskite filling was obtained, that resembles the structures we prepared in the present study using a 10 wt.% perovskite solution (see Fig. 1(c)). Moreover, these results were achieved with electrolyte-based solar cells that also showed a dramatic loss of efficiency over time, ascribed to the perovskite decomposition.[1,15]

The effect of the $TiO_2$ scaffold structure was also investigated and solar cells were fabricated from thicker tube layers. Although these layers showed superior absorbance features (Fig. 2(d)), they provided significantly lower efficiencies compared to shorter (0.4 µm-thick) tubes (Fig. 3(a)). Data in Fig. 3(b) confirm that clearly higher external quantum efficiency (EQE) was measured for thin tube scaffolds. Note that EQE results for devices based on 0.4 µm-thick tubes closely match the corresponding UV-Vis absorption spectrum, this in terms of EQE offset (see Fig. 2(d) and Fig 3(b) for comparison). Solar cells fabricated from 1 and 2 µm-thick tubes showed instead lower cell efficiencies and EQE and, in particular, the latter exhibited unexpectedly lower EQE values in the 600-800 nm range. In other words, thicker tube layers led to promising improvement of light absorption ability but exhibited at the same time a dramatic worsening of the charge collection efficiency – this can be argued also from the $V_{oc}$ and FF data, which significantly decreased with increasing the thickness of the tube scaffolds (Fig. 3(a)), confirming that the rate of charge extraction is limited when using thicker tube electrodes.[3]

Therefore, to further investigate the performance of our solar cells, photo-induced open-circuit voltage decay measurements were carried out.[20] The results (Fig. 3(c)) show that the carrier lifetime tends to decrease as the thickness of the tube scaffold increases. In principle, no significant difference in tube electron transport property is expected to be observed when tube length only slightly varies (*i.e.*, from 0.4 to 2 µm).[13,16,21] These non-ideal results can be ascribed to the non-optimized perovskite deposition into the thicker nanotube scaffolds, meaning that the perovskite structure presents voids and discontinuities (as shown in Fig. 3(d)),[9] so that the hole transport and collection is hindered.

Noteworthy, it was recently shown that 2 µm-thick perovskite-sensitized $TiO_2$ nanotube layers are able to absorb nearly 90% of the incident UV-visible light (and to efficiently convert it into photo-current).[9] Therefore, further optimization of the perovskite deposition technique in order to conformally fill even thicker tube layers is undoubtedly the next challenge to undertake, in view of pushing the solar cell efficiencies of all-solid-state solar cells to higher values.

**Conclusion**

Perovskite-based solar cells were fabricated using anodic $TiO_2$ nanotube layers as blocking layer and electron collector scaffolds. By optimizing a spin-coating deposition, we could obtain interdigitated perovskite structures in intimate contact with the nanotube counterparts, and construct solid-state solar cells without use of hole-transporting material. For solar cells fabricated from 0.4 µm-thick tube scaffolds we reached short circuit current density and efficiency of 19 mA cm$^{-2}$ and 5.0%, respectively. Such relatively high current density is ascribed to the large amount of perovskite material deposited in intimate contact with the nanotube electrode. We envisage a further optimization of the perovskite deposition, along with the use of thicker $TiO_2$ nanotube scaffolds, as the next step to explore in view of fabricating solid state solar cells with enhanced power conversion efficiency.

**Experimental**

Transparent TiO$_2$ nanotube layers were grown by electrochemical anodization of Ti films (thickness of 0.2, 0.5 and 1 μm) that were deposited by electron beam evaporation on FTO glass (TCO22-15, Solaronix). The deposition rate was 0.6 nm min$^{-1}$ at $5 \times 10^{-7}$ - $2 \times 10^{-6}$ mbar (Fraunhofer Institute for Integrated Systems, Erlangen, Germany). Prior to the Ti evaporation, the FTO substrates were cleaned (in acetone, ethanol and de-ionized water, for 15 min each) and coated with a compact layer of TiO$_2$ by spin coating, using titanium di-isopropoxide bis(acetylacetonate) (Aldrich, 75 wt% in isopropanol) dissolved in 1-butanol (Aldrich, 99.8%). The TiO$_2$ compact film enhances the adhesion of the nanotube scaffold to the FTO glass, and also serves as blocking layer.

The anodization experiments to grow TiO$_2$ NT layers were carried out in a two-electrode electrochemical cell with a Pt foil as cathode and the Ti/FTO layers as anode. The anodization voltage was 60 V and the electrolyte was composed of 0.15 mM NH$_4$F (Sigma-Aldrich), 3 vol% deionized (DI) water and ethylene glycol (99.8% purity, <1% H$_2$O; Fluka). After anodization, the TiO$_2$ NT layers were rinsed overnight in ethanol to desorb the electrolyte. Afterwards, the samples were dried in a N$_2$ stream and annealed in air at 450 °C for 1 hour.

The synthesis of the CH$_3$NH$_3$PbI$_3$ light absorber is reported elsewhere.[17] Briefly, CH$_3$NH$_3$I was synthesized by mixing 30 mL of hydriodic acid (37%, Sigma Aldrich) and 27.8 mL of methylamine (40% in methanol, TCI) in an ice bath for 2 h. Subsequently, the solvent was evaporated by placing the solution in an evaporator at 60 °C. The CH$_3$NH$_3$I product was washed with diethyl ether and dried in vacuum. For the synthesis of the CH$_3$NH$_3$PbI$_3$, solution of 10, 20, 40 and 60 wt% of CH$_3$NH$_3$I were mixed to $PbI_2$ (99% Sigma Aldrich) in N,N-dimethylformamine (DMF). To prepare a 10 wt% solution, 0.5 g of $CH_3NH_3I$ and 1.4 g of $PbI_2$ were mixed in 10 ml of solvent.

For the solar cell fabrication, $(CH_3NH_3)PbI_3$ solutions were spin-coated onto the $TiO_2$ NT scaffolds (2000 rpm, 30 s). The samples were then annealed at 100 °C for 20 min. The fabrication of the solar cells was completed by sputtering a Au layer (100 nm) on the top of the device (*i.e.*, onto the perovskite) as counter electrode (see a sketch of the solar cell in Fig. 1 (a)).

The nanostructures were characterized using a field emission scanning electron microscope (Fe-SEM, S4800, Hitachi) and an X-ray diffractometer (XRD, X'pert Philip MPD with a Panalytical X'celerator detector and graphite monochromized CuKα radiation, λ = 1.54056 Å).

UV-Vis absorption spectra were recorded using a PerkinElmer UV-Vis spectrometer with integrating sphere. Photocurrent density-voltage (J-V) measurements were carried out under AM 1.5 illumination provided by a solar simulator (300 W Xe with optical filter, Solarlight) applying an external bias to the cell and measuring the generated photocurrent with a Keithley model 2420 digital source meter. Prior to measurement, the solar simulator was calibrated using a reference solar cell (OPRC22Si-CAL; Optopolymer). The irradiated area of the solar cells was 0.16 $cm^2$ (masking conditions). The External quantum efficiency (EQE) characterization was performed using an EQE Measurement System (Enlitech) equipped with a Xe lamp.

Photo-induced open-circuit voltage decay measurements were carried out using a Autolab PGSTAT 30 Potentiostat/Galvanostat (Ecochemie, The Netherlands) under AM 1.5 illumination provided by a solar simulator (300 W Xe with optical filter, Solarlight).

**Acknowledgements**

This project was funded by the Deanship of Scientific Research (DSR), King Abdulaziz University, under grant no. 16-130-36-HiCi. The authors, therefore, acknowledge with thanks DSR technical and financial support. The authors would also like to acknowledge ERC, Deutsche Forschungsgemeinschaft (DFG) and the Erlangen DFG Cluster of Excellence for financial support.

**Figure Captions**

**Figure 1** - Schematic illustration of the solid-state HTM-free perovskite-based solar cell fabricated from a $TiO_2$ NT scaffold (a). Scanning electron microscopy (SEM) images of $TiO_2$ nanotubes after deposition of $CH_3NH_3PbI_3$ from: 10 wt% solution (b)-(c); 20 wt% solution (d): 40 wt% solution (e)-(g). The inset in (f) shows a high-magnification SEM image of the bottom of a nanotube hosting the perovskite (the scale bar is 100 nm). X-ray diffraction patterns of a crystalline $TiO_2$ NT scaffold before and after deposition of the $CH_3NH_3PbI_3$ perovskite (h).

**Figure 2** - Current density-voltage curves of HTM-free perovskite-based solar cells fabricated from 0.4 μm-thick $TiO_2$ nanotube scaffolds filled with perovskite material by a spin-coating approach (a). Thick (b) and thin (c) perovskite layers deposited over $TiO_2$ NT scaffolds using a 40 wt% solution, by a spin-coating approach carried out at 2000 and 3000 rpm, respectively. UV-Vis spectra of $TiO_2$ NT scaffolds of different thickness sensitized with different amounts of $CH_3NH_3PbI_3$ perovskite (d).

**Figure 3** - Current density-voltage curves along with photovoltaic parameters (a) and EQE spectra (b) of HTM-free perovskite-based solar cell fabricated from $TiO_2$ NT layers of different thickness (the concentration of the spin-coated perovskite solution is 40 wt%). Electron lifetime of HTM-free $CH_3NH_3PbI_3$ perovskite-based solar cells fabricated from tube layers of different thicknesses (c). Cross-sectional SEM image of a 1-μm thick $TiO_2$ tube layer after deposition of $CH_3NH_3PbI_3$ from 40 wt% solution (arrows and dotted area indicate empty space within the NTs) (d).

**Figure 1**

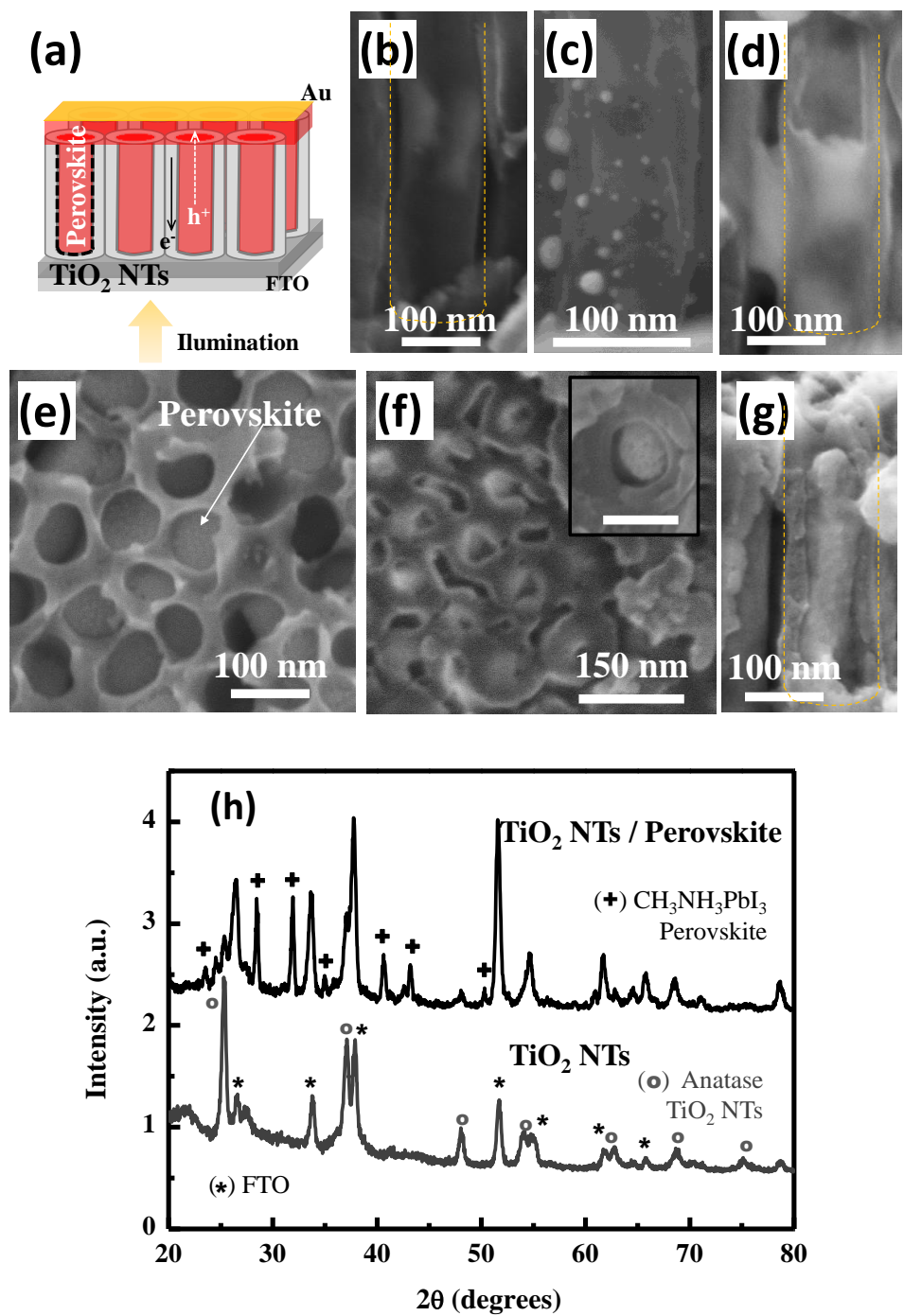

**Figure 2**

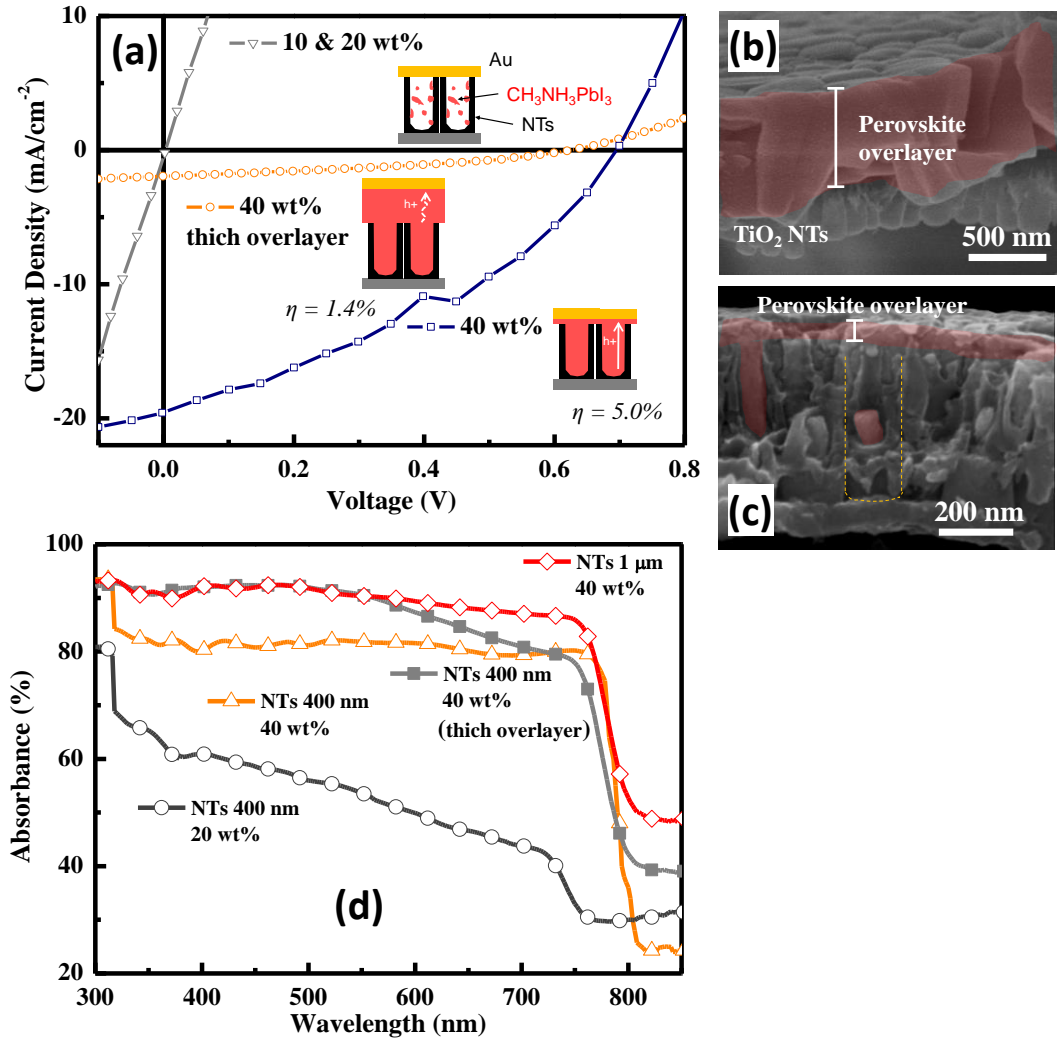

**Figure 3**

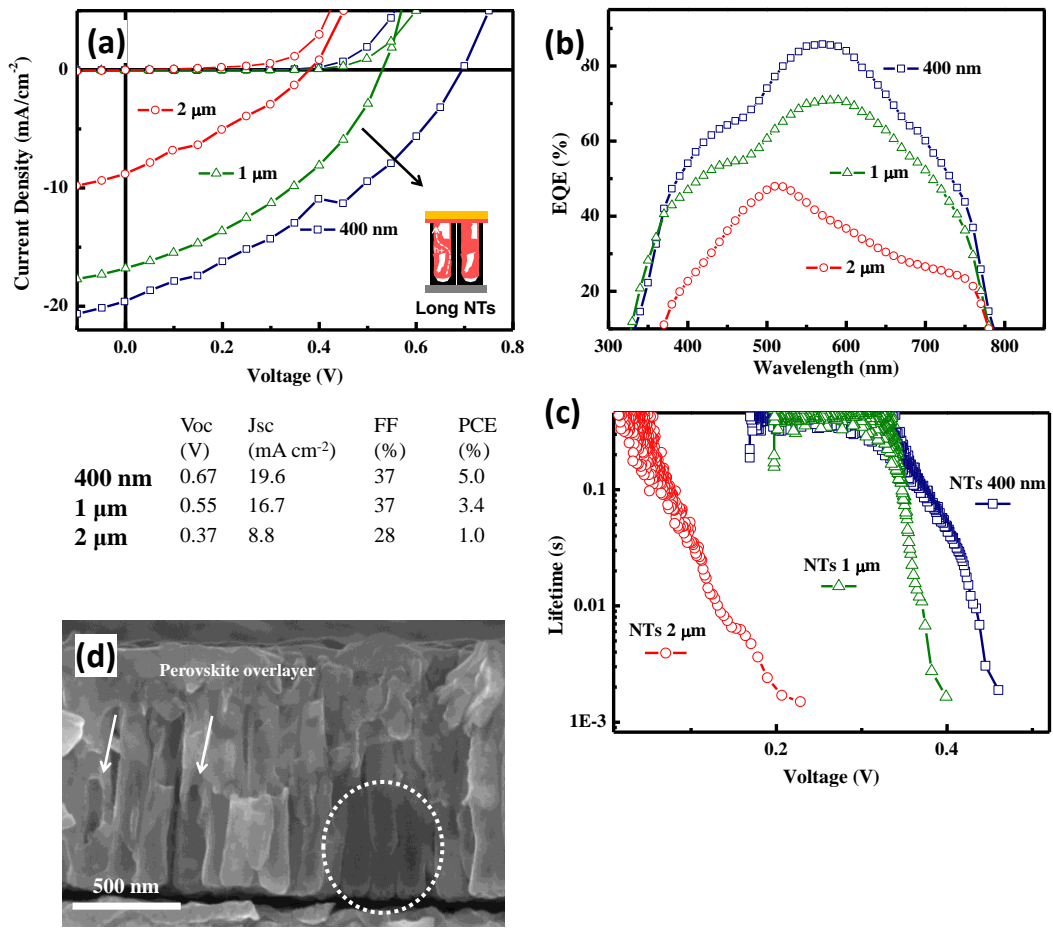

|         | Voc (V) | Jsc (mA cm$^{-2}$) | FF (%) | PCE (%) |
|---------|---------|--------------------|--------|---------|
| 400 nm  | 0.67    | 19.6               | 37     | 5.0     |
| 1 μm    | 0.55    | 16.7               | 37     | 3.4     |
| 2 μm    | 0.37    | 8.8                | 28     | 1.0     |